\documentclass[aps,twocolumn,groupedaddress]{revtex4}
\usepackage{graphicx}
\begin{document}
\title{Continuum model of actin-mediated bacterial propulsion}
\author{V.G.~Benza}
\affiliation{Dipartimento di Fisica e Matematica, Universita' dell'Insubria,
Como, Italy}
\begin{abstract}
The mechanism of propulsion of host bacteria under the action
of actin gel networks
is examined by means of a continuum model of the dynamics of 
the F-actin concentration. The model includes the elasticity
of the network, its attachment
to the host and the polymerization at the interface with it.
A formula for the cruise velocity is derived wherefrom the
respective contributions of elasticity and polymerization
are made explicit. The velocity tends to elongate the
gel in the direction opposite to motion, and in turn
is proportional to the size of its active portion.    
It is shown that the motion can start only past
a finite latency time: the condition for the onset of motion
is explicitly given.It is numerically found that at steady
state the motion has a pulsating behavior, with sudden decelerations
and subsequent recoveries.
\end{abstract}
\maketitle
\vskip2pc
\section{Introduction}

The mechanism of actin-induced propulsion has been
the object of increasingly sophisticated experiments 
in recent years \cite{sootheriot,bernheimprostsykes}.
The phenomenon is of interest as a paradigm of more
complex situations, and exhibits a variety of fascinating
and intriguing features, which keep stimulating new research.
In short, an alien host
(e.g. a bacterium or a polystyrene sphere)
can make the cytoplasmic material to react by
polymerizing the resident G-actin into an F-actin gel network surrounding
the host's surface;
in due time the network evolves into an asymmetric shape,
typically comet-like.  
The host eventually starts moving, and its cruise
can last for relatively long times at almost constant velocity.
We will focus here on two features of the process;
the first one is a threshold mechanism  
at the onset of motion:
the velocity of  the host rises from  zero to its
cruise value quite abruptly, while 
the mass of the gel undergoes a smooth growth.
The second feature is a correlation between the cruise velocity 
 and the length of the comet \cite{sootheriot}:longer tails
generate higher velocities. 
This suggests that the dominant role in propulsion 
stems from a bulk force, rather than from an interface
effect, such as the polymerization rate at
the host's surface. 
In fact in cases of growth over convex surfaces 
it has been observed and proved
that the gel elasticity has a primary role:  
the stress
can inhibit the polymerization at the 
interface \cite{noireauxsykes} and become  
the main propulsive factor.
Disk-shaped
hosts
can as well be put into motion \cite{schwartzmcgrath};
one could argue that since the gel coating 
entirely surrounds
the host, the  convex portions of its surface
are in all cases under the action of stress.
It is difficult though to believe that this
is an exhaustive explanation
when both the cruise velocity and the elongation of the
comets are perpendicular
to the flat faces of the disks.
One must add to the picture
the links connecting the gel to the host \cite{marcysykes};
such links terminate on the surface at
nucleation points , whose distribution is determined by
the combined action of various activator groups \cite{activator}.
The oversimplified digest given above 
actually translates into a rather
intricate dynamical problem,where 
the global 
elastic action of the gel, 
the gel-host links and the polymerization
rate come together into play.
This problem is formulated here in terms of
a one-dimensional model where the system configuration
is described by the gel concentration.
The gel velocity
is self-consistently determined by generalizing
to the time-dependent case the arguments presented in Ref.
\cite{chaikinprost}.
The time evolution of the gel concentration is described
by a deterministic equation, which includes elasticity
and polymerization/depolymerization;
the approach being of mean-field type, stochasticity
is modeled by a diffusion term.
In our formulation
the interaction with the host turns
into the boundary 
condition for the concentration:
the evolution at the boundary
has the form of the tethered ratchet
model of Mogilner and Olster \cite{mogilneroster}.
One of the main results of the paper
is an explicit formula for the velocity (see below Eq. \ref{V}),
wherefrom its correlation with the comet size
becomes transparent.
Furthermore, the origin of 
the threshold
mechanism at the onset of motion is made clear,
as a natural outcome of the model. 
The numerical results
shown here 
exhibit this threshold mechanism as well as
an intermittent behavior of the velocity at steady
state.   
The system starts moving past
a given ``latency'' time, which depends on the rate
of accretion of the gel, the elasticity and the
nucleation rate.
The interface polymerization rate can have some relevance
in the initial stages after start, but at steady
state
the bulk contribution prevails: the resulting motion
has a pulsating behavior, 
with sudden reductions of the velocity and consequent recoveries,
as in a system alternating  ``loading'' phases at low speed
with  ``unloading'' phases at higher speed. 
 (see below Fig.~\ref{fig1}).
\par
\section{Model}

As anticipated, we are going to establish a 
one-dimensional minimal model 
for the dynamics of the actin gel network
during its propulsive action over some host object.
We take a macroscopic point of view, where
the concentration and  the local velocity are 
a natural choice as dynamical variables describing the gel. 
Under fairly reasonable assumptions
one can get rid of the space dependence of the velocity.
The resulting ``average'' velocity is determined
by the concentration: hence
the system is described by this single variable.
In turn
the average velocity, which we identify with 
the velocity of the host, appears in the evolution
equation for the concentration (see below Eq.\ref{EE}).
Let us first show how the host velocity is determined. 
When the concentration $c(x,t)$ has a well defined peak,
say at $x_{max}$, one can approximate the average of the velocity
field $v(x,t)$ as: 
$<v(t)> \approx v(x_{max},t)$.
This obvious remark is not a real progress,
unless one relates the unknown quantity $v(x_{max},t)$
with the value of the velocity field at some other point
$x_{0}$ where it can in principle be determined:
 such a point is  the contact
of the gel with the host \cite{chaikinprost}. 
Indeed on the basis of the standard rate theories \cite{arrhenius}
the velocity at the interface is a function
of the elementary work  $\Delta W$ according with the formula:

\begin{equation}
\label{AR}
v_{A}=v_{+} \exp({- {{\Delta W} \over {k_{B}T}}}) - v_{-}
\end{equation}
where $v_{+},\, v_{-}$ are the free polimerization/depolymerization 
velocities.
One has then:$\,v(x_{0},t)=v_{A}(t)$; the link between $v(x_{0},t)$ 
and the unknown value $v(x_{max},t)$
is provided by
the law of mass conservation.
From now on, unless otherwise stated, we are going to use 
a reference frame comoving with the host, say
at velocity $v$; the mass conservation reads:
\begin{equation}
\label{M}
{d \over dt} c(x,t) + {d \over dx}[(v(x,t) -v)c(x,t)]=0.
\end{equation}
%G.B.:mettere derivate parziali
The role of symmetry breaking in the
onset of motion is not examined here;
let us then assume, e.g., that the gel has grown to the left of the host,
so that when the host moves $x_{0}$ drifts rightwards.
We integrate Eq.\ref{M} over the interval
$x_{b} \le x \le x_{0}$, where $x_{b}$ is a generic position
in the gel, and obtain:
\begin{equation}
\label{REL}
\int_{x_{b}}^{x_{0}} dx {d \over dt} c(x,t)+[v(x_{0},t)-v]c(x_{0},t)=
[v(x_{b},t)-v]c(x_{b},t)
\end{equation}
Let us finally
assume that the gel and the host drift rigidly together, 
i.e. that $v \equiv v(t)=v(x_{max},t)$; 
upon chosing in Eq. \ref{REL} the generic position
at the peak of the concentration $(x_{b}=x_{max})$ we
get the relation we were searching for:  
\begin{equation}
\label{VV0}
v(t)=v(x_{0},t)+{1 \over c(x_{0},t)} \int_{x_{max}}^{x_{0}} dx {d \over dt}c(x,t)
\end{equation}
This formula 
gives the velocity  $v(t)$ as a nonlocal
function of the concentration: the evolution equation 
for $c(x,t)$ will make our treatment self-consistent.
Let us now discuss how such an equation can be
determined.
The dynamical response of a
normal 
gel, where the structure of the network is fixed
by the distribution of crosslinks and of 
flexible chains,
is characterized by 
well defined values  of the coupling constants, such as, e.g.,
the Young modulus in the case of elasticity.
A network of semiflexible, rather than
flexible, chains is apparently more apt to describe the 
actin filaments; whenever its structure is fixed
its elastic response is also defined, but differs
from that of a normal gel \cite{mackintosh}.
Apart from this difference one must be aware of the fact that 
the actin network of the comets has no predetermined structure:
its couplings evolve with the configuration, 
e.g. the elastic response
changes as the mass and crosslink densities
build up or decay: it would be appropriate
to consider both populations.
In the present ``minimal'' context, where we limit ourselves
to the mass density, the compression modulus
depends on $c(x,t)$ and
the role of crosslinks is included in an effective coupling constant
$C$; we refer to the Section Elastic Energy for the
explicit form of the elastic term.

The environmental factors,
such as the "feeding" of polymerization,
the depolymerization, the intrinsic fluctuations of
the concentration at the microscopic scale, can be 
assumed to be proportional to the local concentration.
Sure enough, the environment evolves with 
the structure as well,  but we will disregard its
time evolution for the time being.
We consider
a potential $U(x)$ (a growth rate),describing the
"feeding" and "decaying" processes, in dependence with its sign.
Furthermore, a diffusive term accounts
for the small scale disorder of the environment,
which perturbs a purely deterministic buildup
of the network.
We add a quadratic term describing the short range
self-repulsion between different portions of the structure.

Under such factors, in a static reference frame 
the concentration would then evolve according to
the equation:
\begin{equation}
\label{POP}
{d \over dt} c = U(x-vt)c -b c^{2} +D {d^{2}\over {d^{2}x}} c
\end{equation}
where the time dependent potential accounts 
for the motion of the "activators" which are
are predominantly concentrated around 
the host.
If, e.g., the potential has a constant value
$U>0.$ in the feeding region, its action combined with the
repulsive term $-b c^{2}$ drives $c(x,t)$ towards
the value $\bar c = U/b$. 
The host enforces the condition 
$c(x,t)=0.\, (x>x_{0})$;
furthermore, the gel is attached
to it at the nucleation
points and grows against it when the actin filaments polymerize.
We model this along the lines
proposed by Mogilner and Oster \cite{mogilneroster},
i.e. by means of two concentrations $a(t),w(t)$ 
respectively describing the "attached" and"working" portions
of the gel. Their evolution determines the
boundary condition of  $c$ at the interface:$c(x_{0},t)=a(t)+w(t)$.
We now go over and write the  evolution equation
including the elastic term; this term
is proportional to the deviation of the concentration
with respect to a reference "background" configuration \cite{brgel}, and
is opposite to it.
If, e .g., an excess of mass has accumulated in some region 
during the process of buildup of the network, the elastic force
acts as in a spring compressed in the $x$ direction,
and elongates the excess of mass along $x$ (see the Section
Elastic Energy). 
The frame comoving with the host has coordinates
$x'=x-vt,t'=t$; in these variables the evolution equation
has the following form:  
\begin{eqnarray}
\label{EE}
{d \over dt}c(x,t)& =& D {d^{2} \over {d^{2} x}} c(x,t) +v {d \over dx}c(x,t)\nonumber\\ 
+ U(x) c(x,t) &-&b c^{2}(x,t) -\Gamma {{\delta E_{el}} \over {\delta c(x,t)}} 
\end{eqnarray}
where we have dropped the primes;
$E_{el}$ is the elastic energy and 
$\Gamma$ is the relative decay rate.
Notice that a drift term has appeared, in going to the
moving frame: it  
has the effect of
 elongating the profile of
$c(x,t)$ in the direction opposite to $v$, as if 
the concentration experienced  a backward wind. 
The higher the velocity, the stronger the effect:
the elongation of the profile "measures" the
velocity.
The velocity and
the elongation of the profile actually have a link which goes beyond
this purely hydrodynamic effect.
In fact, according with Eq.\ref{VV0}
the velocity $v$ is in turn $c$-dependent, and
upon inserting the l.h.s. of Eq.\ref{EE} into Eq.\ref{VV0} we finally get:
\begin{eqnarray}
\label{V}
 v={{v(x_{0}) c(x_{0})}\over {c(x_{max})}}&+& {1\over c(x_{max})} \cdot \lbrace D ({{ dc}\over{dx}})_{x_{0}}+ \nonumber\\
\int_{x_{max}}^{x_{0}}dy[U(y)c(y)&-& b c^{2}(y)-\Gamma{{\delta E_{el}}\over{\delta c(y)}}]\rbrace\\
\nonumber
\end{eqnarray}
where $x_{max}<x_{0}$ and the time dependence of the arguments is understood.
In the time independent case only the first term survives,
and the formula simply states the conservation of stationary
fluxes \cite{chaikinprost}; the boundary term $D({{dc} \over {dx}})_{x_{0}}$
is negative, as one easily verifies: it averages
over the region $x_{max}<x<x_{0}$
the loss in propulsion originating from network's disorder.

\section{Numerical Results}
\begin{figure}
\includegraphics[width=3.0in]{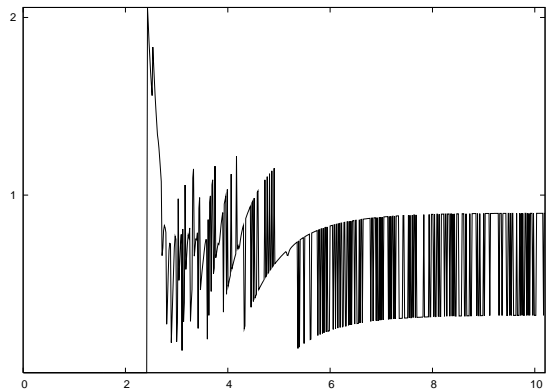}
\caption{Typical behavior of the velocity (in units of
10.\, nm/sec) versus time (in seconds); notice
the discontinuity at the onset of motion, the transient 
before reaching steady state and the intermittent behavior at steady state.
The values of the parameters are reported in the text}
\label{fig1}
\end{figure}

The formula \ref{V} is one of the primary results of the paper;
it shows how
the velocity gets contributions 
both from the interface and from the bulk.
Notice that $v$ goes as the inverse of the maximal concentration;
the elastic force must counter the formation
of exceedingly high concentration peaks 
in order to have higher velocities.
The bulk contribution becomes increasingly relevant
as the peak at $x_{max}$ moves
off  $x_{0}$ in the backward direction.
In the feeding region, close to the interface,
we have $U(x)=U>0.$: the network grows and contributes
positively to the propulsion.
The diffusion constant D $(D=5. 10^{-3}(\mu m)^{2}/sec)$ has a minor effect: 
its value has been chosen two orders of magnitude smaller
than the diffusion constant of monomers within the gel
estimated in Ref.\cite{softlisteria}.
The elastic and self-repulsion terms
operate positively or negatively dependending on
the configuration. They give rise to 
an intermittent behavior of the velocity even at steady state;
typically the velocity 'jumps' between two (or more) values,
staying most of the time in one of them, which can
be identified with the cruise velocity of the host.
Among the temporary, intermittent velocities
we found negative values as well: this happens, e.g., 
when the rate 
$\Gamma$ is half the value $\Gamma =2.\, (pN \cdot sec)^{-1} (nm)^{-2}$
used in the simulation presented here. 
At steady state the motion can be depicted as a sequence of very
short ``charging''steps, where the velocity
is small or negative, followed by longer
advancement steps.
A typical behavior is shown in Fig.~\ref{fig1},where $U=1. (sec)^{-1}$:
the cruise velocity is $v \approx 8.\, nm/sec$,
and on  average $x_{0}-x_{max} =250.\,nm$. 
The velocity is
quite near to  $10. nm/sec$, established
in Ref.\cite{sootheriot} as a minimum value for 
classifying an object as being in motion.
At a closer inspection  
the concentration profile reveals time dependent
small size
deformations tuned with the steps of the velocity. In particular 
the peak position $x_{max}$ evolves with an intermittent behavior:
temporary compressions are subsequently released. 
Let us call active region the portion of space
going from the peak to the interface; in the example
shown in Fig.\ref{fig1} this region goes slightly beyond
the feeding region, which has size $200.\,nm$ (see Fig.\ref{fig3}).
The formula for the velocity Eq. \ref{V} shows that the bulk contribution 
is proportional to the size of the active region.
A linear relation between the length of the comet
and the  cruise velocity has been observed in
very accurate kinetics experiments by Soo and Theriot \cite{sootheriot}: 
we believe that our results are consistent with
such observations.
Actually there are two concomitant effects:
the drift term elongates 
the profile by a factor proportional to $v$,
in turn a more elongated profile generates a larger $v$.
Let us examine the interface contribution. 
According to Eq. \ref{AR}, $v(x_{0})$ 
is a decaying function  of the elementary work $\Delta W$:
\begin{eqnarray}
\label{W}
\Delta W &=& \delta l \cdot F(x_{0})\equiv \delta l \cdot \sigma_{x,x}(x_{0})\nonumber\\
&=&\delta l \cdot c(x_{0}){a\over w}C_{a}\int_{x_{max}}^{x_{0}} dy [c(y)-c(x_{0})] 
\end{eqnarray}
where $\delta l$ is the elementary displacement
having the order of the size of a monomer $\delta l=2.2\, nm$,
$ {a \over w}\,\,$ the ratio of attached
versus working gel and $C_{a}=2.\,(pN \cdot nm)$ the elastic coupling
of the attached portion of the gel.
The above expression follows from the equilibrium condition
between working forces and load \cite{mogilneroster},
the load here being uniquely given by the attached links
(see the Section Elastic Energy).
Let us examine the order of magnitude of the force
at steady state in the case reported in Fig. \ref{fig1}.
There the ratio $a/w$ is close to
one (see Fig. \ref{fig4}), the active region
has size $x_{0}-x_{max}=250.\, nm$, the peak 
is $c(x_{max})\approx 10.\,$ (the concentration has units $(10.\, nm)^{-1}$)
the deviation $c(x_{max})-c(x_{0})$ is of order
$10^{-1}$ so that one has $F(x_{0})\approx 5.\, pN$.
Taking into account that the Boltzmann factor is
$k_{B}T \approx 4.1 pN \cdot nm$ one can verify that
the contribution to $v$ arising  from $v(x_{0})$  is quite
small as compared with the bulk contribution.

Assuming a free polymerization
velocity $v_{+}=10.\, nm/sec$ and $v_{-}=0.$ we always obtained that 
the bulk velocity
has the dominant role in propulsion, with the exception
of the initial transient following the onset of the motion.
Our simulations start with zero concentration.
A given nucleation rate drives the  attached gel $(a(t))$,
its decay activates $w(t)$ \cite{mogilneroster}. 
As $a(t)$ and $w(t)$ start increasing,
the boundary condition $c(x_{0},t)=a(t)+w(t)$
initiates the evolution of $c(x,t)$.
The velocity does not follow smoothly 
the gel growth,
but rather jumps abruptly from zero to a finite value:
the onset of motion has  a threshold.
The motion can start
provided that the gel generates a positive propulsion:
this happens when the maximal concentration
rises above $c(x_{0},t)$: at this point an excess
of mass has been created on the back of the host,
and the elastic response to it translates into motion.
If the gel growth rate  is too small, 
the maximal concentration stays locked at the interface,
and the motion cannot start.

\begin{figure}
\includegraphics[width=3.0in]{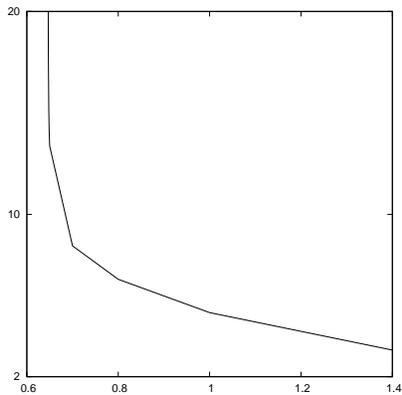}
\caption{Latency time (in secs) versus the
growth rate $U$ (in units $(sec)^{-1}$);the size of the feeding region 
is $x_{f}=200.\,nm$; notice that when $U < \bar U \approx 0.6$ 
the motion cannot start;
these results where obtained with $\Gamma =0.1 \,(nm)^{-2}\cdot (pN sec)^{-1}$,
$C_{a}=1.\,pN \cdot nm$, $C_{w}=0.5\, pN \cdot nm$; all the other parameters 
have values as reported in the text}
\label{fig2}
\end{figure}
If the growth rate is sufficiently high the maximum of $c$ stays at $x_{0}$, 
but only for a finite ``latency'' time.
This time is strongly dependent on
the rate $U$ in the feeding region; in Fig.~\ref{fig2}
it is shown that it diverges as $U$ decreases towards a
finite value $\bar U \approx 0.6$.
Similarly, one can verify that shorter feeding regions generate longer 
latency times. 
As  the peak unlocks from $x_{0}$, the host is instantly put
into motion, and its velocity
$v$ abruptly departs from zero, with a value
depending on the mass accumulated during the
latency.
In the evolution  equation Eq. \ref{EE} the drift term is 
simultaneously activated: the profile
is faced with an abrupt backward wind, and
must adapt its shape to it.
\begin{figure}
\includegraphics[width=3.0in]{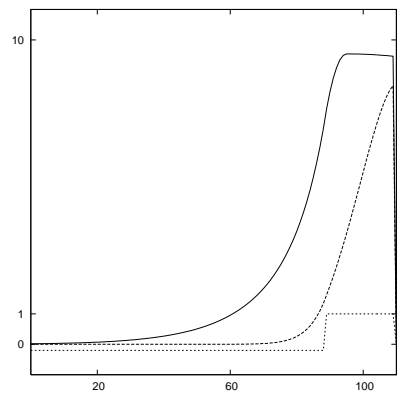}
\caption{From bottom to top: potential U(x); profile c(x) at t=2.sec, 
right before 
the onset of motion (dotted line);
c(x) at steady state (t=15.sec)(continuous line).
The abscissa has space unit $\bar a=10.nm$. The vertical axis is in 
units $(sec)^{-1}$ 
(U(x)), and  $(10.\, nm)^{-1}$ (c(x))}
\label{fig3}
\end{figure} 
A transient follows, during which
the active region modifies its size with strong 
fluctuations, until the system  
reaches a steady state profile.
In Fig. ~\ref{fig3} we compare the profile during 
latency, but close to start, with the steady state
profile; the potential is superimposed [$U(x)=1.$ in the
feeding region, $U(x)=-0.2$ in the decay region].
\begin{figure}
\includegraphics[width=3.0in]{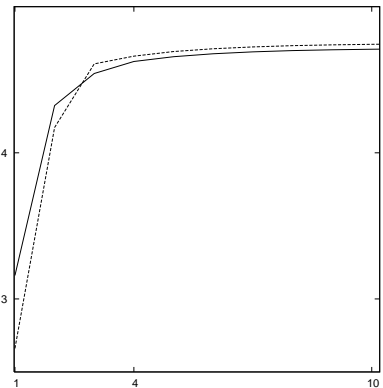}
\caption{Time behavior of the variables a (continuous line) and w 
(dotted line), representing
the attached and working part of the gel at the interface;
same parameters as in Fig. \ref{fig1};the vertical axis has units
$(10.\, nm)^{-1}$, the time unit is $1.\,sec$. At t=1.
$ w$  is still sensibly smaller than $a$. Notice that at some
time 
$t\approx 2.$ $w$ grows above $a$: this event is close to the
onset of motion }
\label{fig4}
\end{figure}
The steady state motion results from 
the combined action of the drift
term and of the elastic forces; the
drift term in particular marks the shape of
the steady state profile, which appears smoothly elongated away
from the interface as compared with its form at the
onset of motion.
A stationary transfer of mass from the feeding region
to the queue  maintains the steady state profile.

\section{Elastic energy}
We write the elastic energy as a quadratic form
\begin{equation}
\label{EL}
E_{el}= {C  \over 2}\int dx \int dy \,\delta c(x,t)\cdot G(x,y)\cdot\delta c(y,t)
\end{equation}
where $\delta c(x,t)= c(x,t) -\bar c(x,t)$ is the deviation
from a "background" configuration $\bar c(x,t)$,  $C$
is an energy whose value must account of
the effect of crosslinks.
The kernel  $G(x,y)$ has the role of bounding 
the range of elastic response within a finite
length $\lambda$: 
$G(x,y)=\exp({-|x-y| \over \lambda})$.
We define  background or reference configuration 
a configuration which is in static
equilibrium but only over a suitable time scale.
New configurations generated at shorter time scales
are attracted towards it.
In the present paper we give a simplified treatment 
where the elastic interaction is limited to 
the active region: we believe that this region
has the dominant role in the propulsion.
In the bulk of the network the elasticity 
operates against changes in the concentration profile.
The background configuration results from the past
hystory of the structure.
For the time being we disregard memory effects and
simply mimic the rigidity of the system with the
self-repulsion term included in Equation \ref{EE};
the relative coupling constant has value $b= 0.5\, nm/sec$.
Let us focus now on the active region:
there a natural choice for the background configuration
is given by the concentration at the interface.
This is because at the host's surface the concentration
is fixed by the boundary condition
$c(x_{0},t)=a(t)+w(t)$, so that in the adjacent region
the profile must be predominantly attracted towards
$c(x_{0},t)$.
In specializing the energy Eq. \ref{EL} to the active region
we let then $\bar c(x,t)=c(x_{0},t)$;it is
understood that the integrations are
extended over the interval $x_{max}<x<x_{0}$,
and we assume $\lambda>>(x_{0}-x_{max})$.
About the form of the coupling $C$,
one must consider 
that the attached and working portions operate
in opposite ways and with different strength. 
The variable $a(t)$
is related with links connecting the gel to the
host: their role is of the same nature as that of the 
gel-gel crosslinks, which determine the bulk's elastic
response.
Hence $a(t)$ attracts the system towards the background
configuration while $w(t)$ acts in the opposite direction.
The specific form of the coupling $C$ must then be:
$C= {a\over c(x_{0})} C_{a} - {w \over c(x_{0})} C_{w}$ 
where with $C_{a}$ and $C_{w}$ we have separated the 
elastic response of the
attached links from that of the working filaments
(we always assume that $C_{a} > C_{w}$; in particular in
the simulations presented here we have $C_{a}=2. C_{w}; 
C_{w}=1.\,pN \cdot nm)$).  
As we assumed that the length $\lambda$ is much larger than the
active region, we average over the space dependence of the
deviation:
$\delta c(x,t) \approx \delta c \equiv {1\over2}( c(x_{max},t)-c(x_{0},t))$.
The total elastic force, proportional to $\delta c$, 
stabilizes the gel provided that the contribution from
the gel-host links is the dominant one, i.e. under the condition
$C_{a}\cdot a -C_{w}\cdot w > 0.$
Let us add some comment on the dynamics of $a(t)$ and $w(t)$:
 their evolution equations  
are slightly modified with respect to  equations (1) and (2) 
of Ref. \cite{mogilneroster}. Here $w(t)$ gets a contribution not only 
from the detached links, but also from the
adjacent gel $(\bar c\equiv c(\bar x);\,\, (x_{0}- \bar x)<<1)$:
\begin{eqnarray}
\label{MO}
\dot a &=& j  - \delta(v) a\nonumber\\
\dot w &=& \delta(v) a -d_{c} w +D (\bar c -2\,w)
\end{eqnarray}
here $\delta(v)=\delta_{0}(1.+ {v \over {\bar v}})$
is the velocity-dependent detachment rate, $d_{c}$
is the capping rate and $j$ is the nucleation rate
per space unit.  
We assigned to these variables the following
values:$j=0.5\, (nm \cdot sec)^{-1}$ $\delta_{0}=1.\, (sec)^{-1}$
$\bar v = 100.\, nm/sec$, $d_{c}=0.8\, (sec)^{-1}$.
It is necessary to connect this
formalism with the usual treatment of
elasticity: this is done by assuming  
that the system is incompressible, i.e. that the elementary mass
contained in a generic interval $\Delta x$ is conserved
under deformation: $\delta (c(x) \Delta x)=0.$
The condition above gives:
\begin{equation} 
\delta c(x) = - c(x) {d \over dx} \Delta x = - c(x) u_{x,x}(x)
\end{equation}
where $u_{x,x}(x)$ is the strain tensor in one dimension.
Eq. \ref{EL} then reads:
\begin{equation}
E_{el}= {C\over 2} \int dx \int dy [c(x) u_{x,x}(x)]G(x,y) [c(y) u_{y,y}(y)]
\end{equation}
this should be compared with the standard form of the elastic
energy:
\begin{equation}
E_{el}=B \int dx (u_{x,x}(x))^{2}
\end{equation}

The compression modulus $B$, which in one dimension 
is an  
energy/length, in our case is explicitly concentration-dependent.
In fact, 
the stress tensor $\sigma_{x,x}(x)$, which here has the dimension
of a force, is given by 
\begin{eqnarray}
\sigma_{x,x}(x)&=& {{\delta E_{el}}\over {\delta u_{x,x}(x)}}\nonumber\\
&=&
c(x){{\delta E_{el}}\over{\delta c(x)}}\approx C \lambda_{eff} c^{2}(x) u_{x,x}(x)
\end{eqnarray}
where $\lambda_{eff}$ is the effective range of the interaction.
We have then:
$B \to  C \lambda_{eff} c^{2}(x)$.
In the case of Fig.\ref{fig3} one would have 
$B \approx C_{a}\cdot (x_{0}-x_{max})\cdot c^{2}(x_{max})\approx 0.5\, nN.$
Correspondingly the elastic stiffness can be estimated as
the ratio $k_{el}=B/(x_{0}-x_{max})$: 
we obtain then $k_{el}= 10^{-3}\, nN/nm$, quite small as compared
with the value $k_{el}= 0.17\, nN/nm$,
reported in Ref. \cite{theriotelastic}.

The value of $C_{a}$ was 
chosen by requiring a
  $\Delta W$ of order $k_{B}T$.
In the equation of motion the elastic coupling $C$
multiplies $\Gamma$, whose value was chosen in order
to make the elastic term strong enough to counter
an unlimited growth in the feeding region.

From Eq.\ref{W} one can argue that a steady state with a ratio 
$a/w << 1.$ would allow for
higher values of $C_{a}$ and consequently would give more
realistic values of stiffness;
the rate $\Gamma$ should be accordingly reduced, in order 
to keep the same strength of the elastic term in the equation of motion.

\section{Conclusions}

We presented a continuum model of actin-mediated
bacterial propagation.The model is a minimal one, in that
it describes the system in terms of a single variable:
the concentration $c(x,t)$.
The time evolution of $c(x,t)$ includes
the interaction with the environment and with 
the host as well as the elastic
interaction in the region adjacent to the host's
surface.
A formula for the velocity is obtained,
where the interface and bulk contributions are made explicit.
This formula, together with the hydrodynamic effect
acting on the profile, makes the correlation 
between the velocity and the length of the actin network explicit:
this result agrees with the behavior observed in Ref. \cite{sootheriot},
and indicates that the bulk, rather than the interface, has the major role 
in propulsion.
The model furthermore motivates on theoretical grounds the existence 
of an intrinsic threshold mechanism underlying the onset of the motion.
A relevant feature of a minimal model is that in the 
numerical simulations
the number of adjustable parameters 
is accordingly quite small. The key parameters
regulating the dynamics of the concentration in the bulk
are the growth rate $U$,the self-repulsion coupling $b$, the elastic
coupling  $C$ and the rate $\Gamma$; the interface
dynamics is controlled by the nucleation, detachment and
capping rates $(j,\,\delta(v),\,d_{c})$.
Preliminary numerical results exhibit the threshold 
mechanism
and its dependence on the growth rate (see Figs.\ref{fig1}, \ref{fig2}).
The ``cruising'' steady state  is characterized by a
pulsating behavior, where the motion undergoes sudden
reductions and subsequent recoveries.\\

Acknowledgements.

The Author wishes to thank M.Cosentino Lagomarsino an B.Bassetti,
who got him acquainted with comets and listeria and encouraged the
accomplishment of this work.


\begin{thebibliography}{1}

\bibitem{sootheriot}
Soo F S and Theriot J A 2005 Large-scale quantitative analysis of sources
of variation in the actin polymerization-based movement of 
{\it Listeria monocytogenes}\, 
{\it Biophys.J.} {\bf 89} 703-723

\bibitem{bernheimprostsykes}
Bernheim-Groswasser A, Prost J and Sykes C 2005 
Mechanism of actin-based motility: a dynamic state diagram 
{\it Biophys. J.} {\bf 89} 1411-1419
 

\bibitem{noireauxsykes}
Noireaux V , Golsteyn R M, Friederich E,Prost J,
Antony C, Louvard D and Sykes C 2000
Growing an actin gel on spherical surfaces {\it Biophys. J.} {\bf 278} 1643-1654\\


\bibitem{schwartzmcgrath}
Schwartz I M,Ehrenberg M,Bindschadler M and McGrath J L 2004
The role of substrate curvature in actin-based pushing forces
{\it Curr.Biol.} {\bf 14} 1094-1098 \\


\bibitem{marcysykes}
Marcy Y,Prost J, Carlier M F, Sykes C 2004 Forces generated during
actin-based propulsion: a direct measurement by micromanipulation
{\it Proc.Natl.Acad.Sci.USA} {\bf101} 5992-5997 \\


\bibitem{activator}
Welch M D, Rosenblatt J, Skoble J, Portnoy D A and
Mitchison T J 1998 Interaction of human Arp2/3 complex
and the Listeria monocytogenes ActA protein in actin filament nucleation 
{\it Science} {\bf 281} 105-108  \\
Cameron L A ,Svitkina T M,Vignjevic D,Theriot J A and
Borisy G G 2001 Dendritic organization of actin comet tails 
{\it Curr.Biol.} {\bf 11} 130-135\\

\bibitem{chaikinprost}
Gerbal F,Chaikin P,Rabin Y and Prost J 2000
An elastic analysis of {it Listeria monocytogenes} propulsion
{\it Biophys. J.} {\bf 79} 2259-2275\\


\bibitem{mogilneroster}
Mogilner A and Oster G 2003 Force generation by actin polymerization II:
the elastic ratchet and tethered filaments {\it Biophys. J.} {\bf 84}, 1591-1605\\


\bibitem{arrhenius}
Hill T L 1987 {\it Linear aggregation theory in cell biology} (New York: Springer-Verlag)\\

\bibitem{mackintosh}
Head D A ,Levine A J and MacKintosh F C 2003 
Deformation of cross-linked semiflexible polymer networks
{\it Phys.Rev.Lett.} {\bf 91} 108102-4\\

\bibitem{brgel}
Read D J,Brereton M G and Mcleish T C B 1995 Theory of the order-disorder
phase transition in cross-linked polymer blends
{\it J.Phys.II} France {\bf 5} 1679-1705\\

\bibitem{softlisteria}
Boukellal H,Campa\'s O,Joanny J F,Prost J and Sykes C 2004
Soft Listeria:actin-based propulsion of liquid drops
{\it Phys.Rev.} {\bf E69} 061906-1 \\

\bibitem{theriotelastic}
Parekh S H,Chaudhuri O,Theriot J A and Fletcher D A 2005
Loading history determines the velocity of actin-network growth
{\it Nature Cell Biology} {\bf 7} 1219-1223\\
%in dell:/gel/theriot05.pdf

\end{thebibliography}
\end{document}